# Microscopic Intricacies of Self-Healing in Halide Perovskite-Charge Transport Layer Heterostructures

Tejmani Behera,[a] Boris Louis,[a] Lukas Paesen,[a] Roel Vanden Brande,[a] Koki Asano,[b] Martin Vacha,[b] Maarten Roeffaers,[a] Elke Debroye,[a] Johan Hofkens,[a,d*] Sudipta Seth[a*]

[a.] Department of Chemistry, KU Leuven, Leuven 3001, Belgium
[b.] Department of Materials Science and Engineering, Tokyo Institute of Technology, Tokyo 152-8552, Japan
[c.] School of Materials and Chemical Technology, Institute of Science Tokyo, Tokyo 152-8552, Japan
[d.] Max Planck Institute for Polymer Research, Mainz 55128, Germany

**Abstract**

The stability and performance of halide perovskite photovoltaic devices are critically limited by progressive defect generation and associated local non-radiative losses during operation. Self-healing of defects provides a promising pathway to prolong device functionality, yet the underlying microscopic mechanisms remain poorly understood, particularly the role of interfacial chemistry on trap dynamics and healing kinetics. Here, we elucidate self-healing and defect evolution in triple-cation mixed halide (TCMH) perovskite films and their device-relevant charge transport layer heterostructures subjected to photo-induced damage. Using correlation clustering imaging (CLIM), our recently developed local functional imaging tool, we map spatiotemporal photoluminescence heterogeneity to track defect dynamics in pristine and heterostructure films. The defect healing follows bi-phasic kinetics, with an initial electronic relaxation (tens of minutes) and a subsequent slower phase (~ hours) associated to ionic and lattice rearrangement. Most importantly, our results demonstrate that the chemical nature of charge-transport layers modulates trap activity, healing kinetics, and halide redistribution, with heterostructures exhibiting faster recovery than pristine films, a boon for device resilience. These findings provide new insights into the dynamic interaction between defects, interfaces, and ion migration, and establish a framework for rational design of durable, next-generation perovskite optoelectronic devices.

**Introduction**

Metal halide perovskites (MHPs) combine high power conversion efficiencies in solar cells surpassing 26 %,[1] with the advantages of low cost solution processing,[2] while also demonstrating remarkable versatility across light-emitting diodes (LEDs),[3] lasing,[4] and detector applications.[5,6] Yet instability of halide perovskite optoelectronic devices remains a bottleneck for long-term operation, largely due to persistent defect formation and associated non-radiative (NR) losses. However, unlike conventional semiconductors, MHPs possess an inherent self-healing (SH) ability that allows spontaneous repair of photo-/radiation-, electrical-, and mechanical stress-induced defects without external intervention.[7–14] The unusual self-healing property presents a pathway toward resilient, low-maintenance devices, yet its practical realization hinges on deciphering and regulating the microscopic origins of SH.

SH stems from soft, dynamic, and defect tolerant lattices,[15,16] where many intrinsic point defects are thermodynamically unstable,[17] enabling recovery on timescales from minutes to hours.[8,14] However, SH involves various mechanisms, involving activation-deactivation of metastable traps,[14] transient chemical intermediates,[8] polaron formation,[11] highlighting the complex interplay of electronic-ionic processes.[10,18] Grain boundaries and microstructure further modulate these dynamics.[7,8,18] Despite this progress, understanding on how interfaces, the backbone of any functional device, affect these SH processes remain poorly explored.

The chemical nature of charge transport layers (CTLs) is known to strongly influence defect densities, ion migration, and recombination kinetics.[19,20] While SH has been observed in completely stacked devices,[11,12,21] systematic studies isolating the contribution of individual CTLs are still limited.[22–24] Moreover, most prior work primarily relies on spatially-averaged probes that obscure the heterogeneous, local nature of defect dynamics,[11,12,14,25] possibly due to absence of a methodology that can simultaneously resolve both spatial and temporal evolution of defects. In contrast, recent photoluminescence (PL) imaging studies have revealed pronounced PL fluctuations (blinking) across MHP crystals and thin films,[26–28] arising from intermittent formation of highly efficient yet metastable quenchers ($Q$s).[29,30] A single $Q$ can dominate NR recombination across several microns via long-range energy migration,[31] manifesting as spatially correlated intermittency events.[28,32,33] These fluctuations serve as optical fingerprints of defect generation and subsequent recovery,[27] directly establishing mechanistic connection amongst PL dynamics, defect metastability, and SH, governed by common microscopic processes. Therefore, analyzing spatiotemporal PL fluctuations with advanced analytical methods can provide a robust approach to elucidate defect evolution and healing dynamics during degradation and recovery.

Here, we systematically investigate SH and defect dynamics in triple-cation mixed halide (TCMH) [Cs$_{0.05}$ (MA$_{0.17}$FA$_{0.83}$)$_{0.95}$ Pb(I$_{0.83}$Br$_{0.17}$)$_3$] perovskite films and their heterostructures with different CTLs. By employing correlation clustering imaging (CLIM),[28] our recently developed spatiotemporal functional PL mapping approach, we resolve defect population and their evolution during recovery. We show that interfacial chemistry critically governs transient defect activity, halide redistribution, and healing kinetics. Our results demonstrate that conductive, stable CTLs accelerate recovery, and suppress halide migration, while reactive interfaces exacerbate the defect activity and spectral instabilities. These insights establish a mechanistic connection between defects, ions, and interfaces, offering design principles for durable perovskite optoelectronics.

**Results**

**Evolution of transient defects and PL recovery in pristine TCMH films.** We fabricated pristine TCMH thin films following a previously reported solution processed method (see **Materials and Methods**, SI),[34] yielding uniform surface coverage with typical grain sizes of ~ 200 nm (**Figure S1-S2**). Despite the uniform surface coverage, PL imaging (**Figure 1a (i)**) revealed significant variations in local emissivity. Time-resolved imaging further demonstrates that these spatial disparities are dynamic with considerable local emission fluctuations, exemplified by intensity traces from three distinct regions of interest (ROI 1-3) over 15 seconds (**Figure 1d, Figure S3**). The PL intermittently switches across multiple intensity levels on timescales of tens to hundreds of milliseconds.[35] Over the years, such fluctuations has been observed in various MHPs structures and compositions,[28,35] many of which exhibit spatially synchronous intermittency over extended regions. These spatio-temporally correlated PL fluctuations is generally attributed to stochastic formation and annihilation of metastable NR defects (quenchers ($Q$s) or super-traps),[28,35] which can quench carriers with communication up to several microns.[32] This interplay of defects and carriers, therefore, renders PL fluctuations a direct probe of local photophysical dynamics in MHPs. To quantitatively analyse these fluctuations, we employed our recently developed CLIM methodology, which computes pixel-wise Pearson cross-correlations of PL intermittency time traces with neighbouring pixels, generating a correlation map.[29] Locations dominated by $Q$s manifest synchronous PL fluctuations and register higher correlation coefficients; while lower values indicate mixed or weak fluctuations. CLIM further segments films into functional clusters (cluster map), where each cluster consists of pixels exhibiting correlation values above a defined threshold, corresponding to an area influenced by regional NR defects interacting with a common charge cloud. We leverage this framework to investigate the spatial distribution and evolution of defects, hence clusters, before and after photodamage, as well as during dark healing.

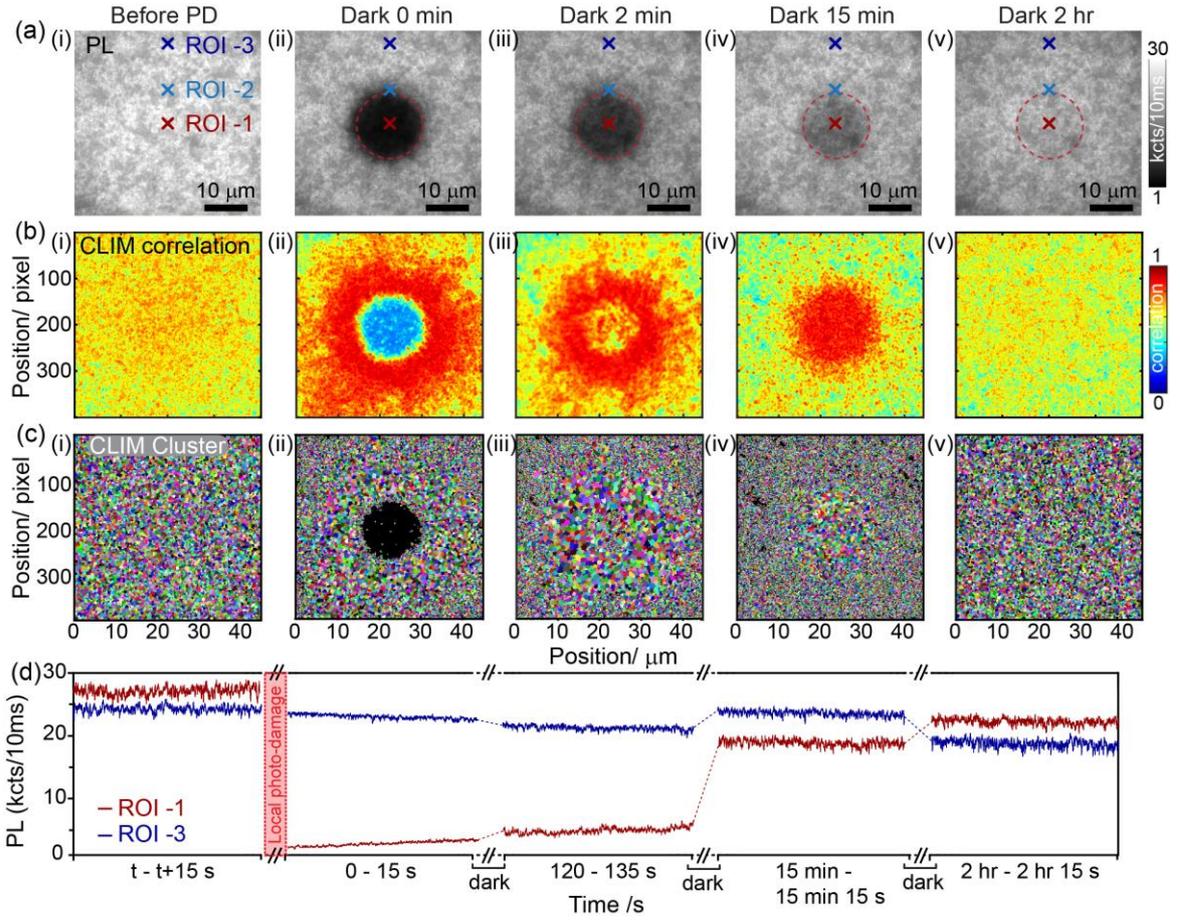

**Figure 1.** Spatial evolution of TCMH film properties during dark recovery: Time sequenced (a) PL images, corresponding (b) CLIM correlation maps, (c) CLIM cluster maps of same area at *(i)* pre-photodamage, and *(ii-v)* post photodamage after 0, 2, 15, 120 minutes of dark healing, respectively. (d) Representative PL traces of ROI -1 and -3. 5x5 pixel$^2$ areas marked in (a) illustrate temporal evolution of emission intensity and fluctuation dynamics both within and outside photodamaged region at different healing stages. Red dashed circle in (a) marks the laser-irradiated area during photodamage.

The CLIM analyses on pristine TCMH film produced an inhomogeneous CLIM correlation map with spatial mean of correlation value ~ 0.6 (**Figure 1b (i)**), indicating significant variability in transient defect dynamics over space. Notably, certain localized regions exhibited higher CLIM correlation values (> 0.85). The corresponding cluster map (**Figure 1c (i)**) captures these domains, identifying areas influenced by common trap populations. To probe spatially resolved SH behaviour, we selectively induced local photodamage (details in **Methods**, SI) and monitored PL recovery during subsequent dark healing. Photodamage led to significant decrease in emissivity within the irradiated area, visible as dark circular zone at the centre (**Figure 1a (ii)**), with nearly three orders of magnitude reduction in PL intensity at ROI-1 (**Figure 1d**, $t$ to $t + 15$ s). Fluctuation amplitudes were substantially suppressed within the damaged zone. Adjacent ROI-2 exhibited an order of magnitude decrease in PL (**Figure S3**), while ROI-3 showed marginal changes in intensity with reduction in fluctuation amplitudes. These observations indicate formation of a high density of NR traps within the irradiated region, responsible for suppressed emissivity and PL intermittency. During dark recovery, the irradiated region gradually regained PL emissivity and fluctuation amplitude (**Figure 1(a, d)**). Partial intensity recovery was observed within the initial few minutes (**Figure 1a (iii), Figure 1d**, 120-135 s), while extended healing over ~ two hours progressively restored the fluctuation dynamics and spatial emissivity close to their pre-damage states. However, full recovery of the mean PL intensity to the original baseline required ~ 24 hours (**Figure S3**).

To further elucidate photophysical changes associated with defect evolution, we performed CLIM analyses during recovery phases. This revealed a characteristic 'donut-shaped' spatial CLIM correlation pattern immediately after photodamage, with significantly reduced correlation value (~ 0.3) within the irradiated area, accompanied by increased correlation (> 0.9) in surrounding periphery (**Figure 1b (ii)**). Cluster maps exhibited enlarged cluster size surrounding the damaged zone relative to pristine film (**Figure 1c (ii)**), indicating that photo-induced defect activity extends radially well beyond the irradiated zone up to ~ 10 μm. As healing progressed, both CLIM correlation and cluster sizes in central region gradually increased, while those in peripheral regions decreased (**Figure 1b, c (iii-iv), Figure S4**). The near-complete restoration of pre-damage CLIM correlation and cluster morphology occurred over ~ 2 hours (**Figure 1b, c (v)**). These results demonstrate that local photodamage triggers mesoscale redistribution of defects, and establish CLIM correlation and cluster mapping as sensitive, spatial probes of dynamic defect densities and behaviors during perovskite photo-darkening and self-healing.

**Mechanistic insights via numerical simulations.** To complement our experimental results and understand how metastable defect density influences spatial CLIM correlation dynamics, we performed Monte-Carlo simulations designed to replicate the spatially heterogeneous PL fluctuations revealed by experiments. Using a uniformly emissive lattice model of size 0.1×6.8×6.8 μm$^3$ (1×50×50 pixel$^3$) without grain boundaries (**Figure 2a**), we randomly distributed the ***Q***s at varying densities ($n_Q$) within the lattice, as detailed in the **Method section** (SI). Each ***Q*** was modelled as an independent entity exhibiting distinct activation–deactivation kinetics, resulting in multi-state PL fluctuations. The quenching capacity of each ***Q*** was constrained to 5% of the total carrier population, estimated from the relative amplitude of fluctuations to the mean (baseline) intensity in an experimentally observed PL trace in pristine TCMH films (**Figure S5** and related discussion, SI).

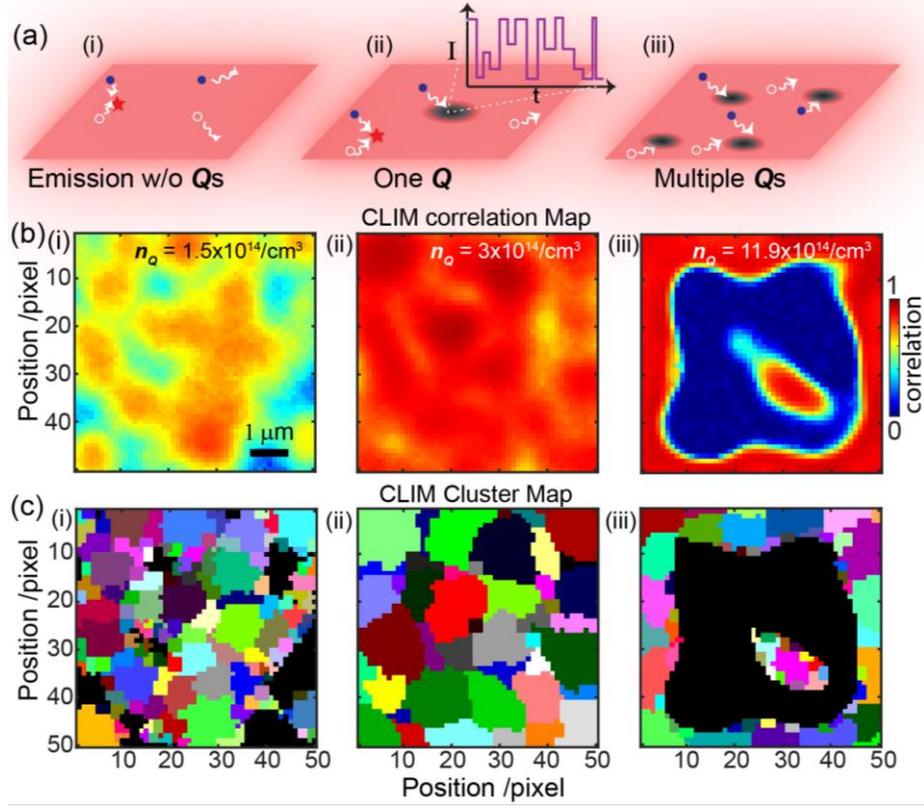

**Figure 2.** (a) Schematic illustration of emissivity modulation within a lattice in presence of *(i)* without, *(ii)* single, *(iii)* multiple independent non-interacting quenchers (***Q***s). Each ***Q*** exhibits distinct multi-level PL intermittency (activation-deactivation) behavior, and influences carrier recombination within certain spatial region. (b) Simulated CLIM correlation maps, and (c) corresponding cluster maps for lattice with dimension 0.1×6.8×6.8 μm$^3$ (1×50×50 pixel$^3$) with ***Q*** density ($n_Q$) of *(i)* 1.5, *(ii)* 3, *(iii)* 11.9 (×10$^{14}$/cm$^3$), reveal the influence of ***Q*** populations on spatial emission patterns.

Simulations revealed that a low trap density ($n_Q$ = 0.7×10$^{14}$/cm$^3$) produced weak spatial CLIM correlation. Increasing $n_Q$ to an intermediate density (1.5×10$^{14}$/cm$^3$) led to localized emergence of spatially distinct, highly correlated domains, which closely resemble the clusters observed in pristine films (**Figure 2b, c (i)**). At $n_Q$ of 3×10$^{14}$/cm$^3$, CLIM correlation across the lattice is enhanced with enlarged average cluster sizes (**Figure 2b, c (ii)**). This mirrored the experimental observations of large clusters at the periphery during early SH stages (0 - 2 min), and within irradiated regions after 15 min of recovery (**Figure 1c (ii-iv)**). However, at high densities ($n_Q \geq 11 \times 10^{14}/\text{cm}^3$), simulations exhibited pronounced PL quenching and severe suppression of fluctuations (**Figure S6**). This led to a collapse in spatial CLIM correlation and disappearance of clusters (**Figure 2b, c (iii)**), consistent with experimental data from irradiated regions immediately after photodamage (**Figure 1b, c (ii)**).

These simulations demonstrate that defect density regulates not only the overall emissivity but also mesoscale emission dynamics. Moderate trap densities foster enhanced spatial CLIM correlation and cluster formation, while excessive defect populations cause severe PL quenching, reduced fluctuation amplitudes, and loss of CLIM correlation signals. The reported defect densities in TCMH single crystal and polycrystalline film are ~ $10^9$ cm$^{-2}$ and $10^{14} - 10^{16}$ cm$^{-3}$, respectively.[36,37] It is worth mentioning that the model assumes uniform maximum quenching capacity for all ***Q***s, whereas in reality, quenching efficiency likely varies depending on defect origin, chemical nature, and local environment. These simulations provide only a qualitative framework to interpret the experimental observations; they do not directly quantify the absolute number of ***Q***s in TCMH films. Nevertheless, despite its simplified nature, these simulations effectively reproduced key experimental trends and delivered mechanistic insights into SH phenomena in MHP films.

**Impact of charge transport layers on defect distributions and PL dynamics.** To bridge the gap between intrinsic film behavior and their device-relevant architectures, we next investigated perovskite/CTL heterostructures. Such interfaces are not passive boundaries; they reshape local energetics,[38] ion migration,[39] and strain fields.[40,41]. Probing these interfacial effects is essential to understand defect dynamics and healing processes, which ultimately dictate long-term device stability and performance.

To address this, we systematically analyzed the spatiotemporal evolution of PL and SH in TCMH films on various electron transport layers (SnO$_2$, ZnO), and hole transport layers (PEDOT:PSS, PTAA). The SnO$_2$ layer forms adhesive, benign contacts via Sn-halide bridges, Pb-O linkages, and H-bonding with perovskite that passivate interfacial defects.[42] In contrast, basic nature of ZnO deprotonates A-site cations of perovskite forming –OH or ZnOH, and Zn-halide species that lower reaction barriers and increase traps, triggering interfacial degradations.[43,44] On other hand, sulfonates of PEDOT:PSS engages in ionic interactions and H-bonding with A-site organic cations and PEDOT–halide interactions; however, acidic and hygroscopic PSS drives corrosion and moisture uptake, destabilizing buried interface with perovskite.[45–47] The hydrophobic and electronically inert nature of PTAA provides a low-trap interface. The alteration in surface interactions and polarity marginally change the surface morphology and crystallographic orientations of TCMH films(**Figure S1, S7**).

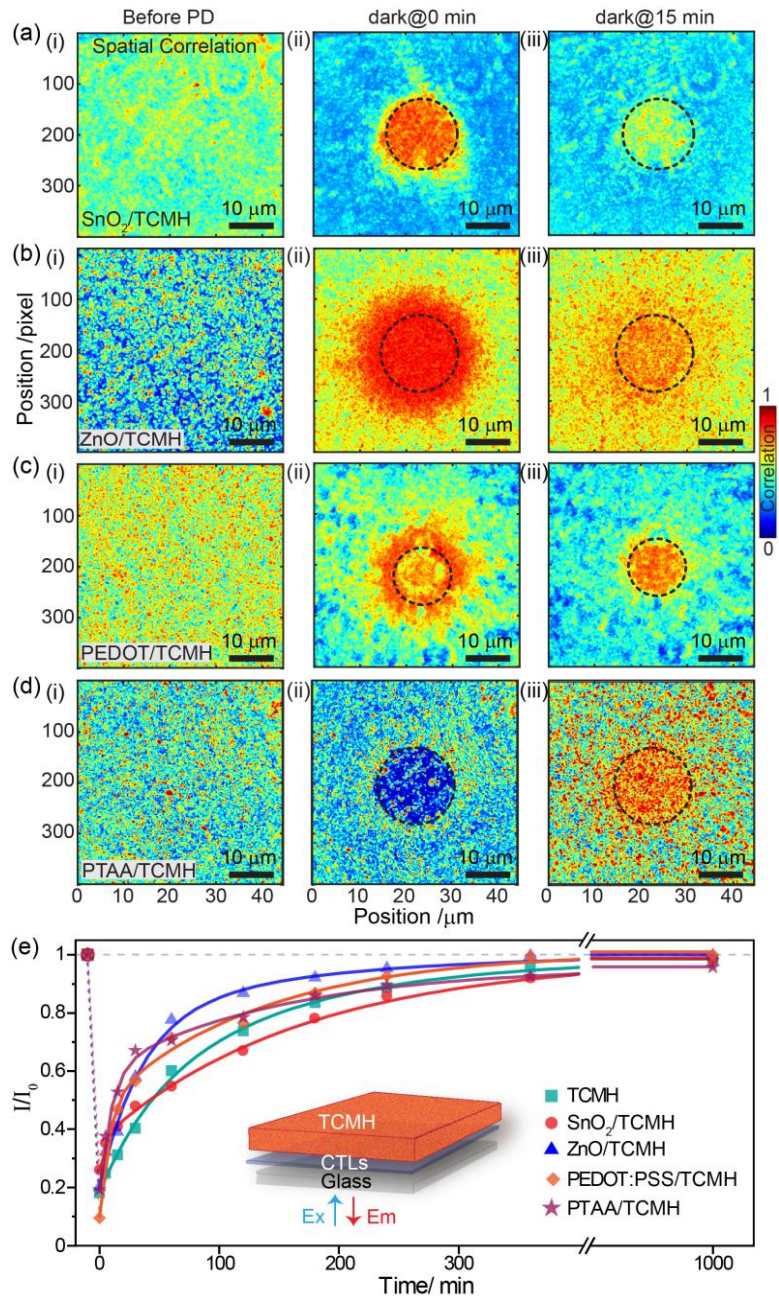

**Figure 3.** Evolution of CLIM correlation during self-healing in TCMH hetero-structure films with (a) $SnO_2$, (b) ZnO, (c) PEDOT:PSS, (d) PTAA, shown for *(i)* pre-photodamage, and after photodamage with recovery of *(ii)* 0 min, *(iii)* 15 min. (e) Normalized intensity evolution of photodamaged region (ROI ~ 2 x2 µm$^2$) for pristine TCMH (cyan), and interfaced with $SnO_2$ (red), ZnO (blue), PEDOT:PSS (orange), and PTAA (purple). Intensities are normalized to pre-photodamage baseline; solid lines are bi-exponential fits to intensities during self-healing.

SnO$_2$/TCMH samples exhibited reduced CLIM correlation (spatial mean ~ 0.53 ± 0.06) and lower emissivity compared to pristine films (~ 0.6 ± 0.07) (**Figure 3a (i), Figure S8**). The PL fluctuations traces of these heterostructures are provided in **Figure S9**. In case of ZnO/TCMH films, spatial CLIM correlation was even more heterogeneous with broad distribution with a lower spatial mean of 0.39 ± 0.18 (**Figure 3b (i)**). This is indicative of higher trap density at ZnO interface. Moreover, PEDOT:PSS/TCMH heterostructures showed a broader distribution of CLIM correlation values (spatial mean ~ 0.56 ± 0.12) with the lowest emissivity, suggesting pronounced mesoscale activation-deactivation dynamics of local NR centers. Conversely, films with PTAA displayed lower spatial average CLIM correlation (~ 0.44 ± 0.17) yet higher PL intensity compared to PEDOT:PSS, consistent with a previous report.[48] CLIM cluster maps further demonstrate that SnO$_2$ and PEDOT:PSS interfaces favored formation of spatially continuously dispersed clusters, indicating relatively homogeneous defect activity across these films. In contrast, ZnO and PTAA based heterostructures exhibited fragmented, spatially localized clusters indicative of enhanced local defect activity at interfaces (**Figure S10).** Given high carrier diffusivity, even localized interfacial traps can influence mesoscale optoelectronic properties. The change in interface interactions modulates the defect density that led to significant variations in spatial CLIM correlation and clusters in heterostructures compared to films on glass (**Figure 3 a-d (i)**). However, the spatially averaged emission and CLIM correlation values on heterostructures were comparable or fell below glass, consistent with the interface driven charge separation and transient defect modulation that supress radiaitve recombination (**Figure S8**).

**Interfacial influence on self-healing kinetics.** To assess SH dynamics across interfaces, we subjected the heterostructures to analogous localized photodamage and monitored recovery. All CTL-interfaced films exhibited localized PL quenching and increased PL intermittency within damaged areas (**Figure 3 (a–d) ii**). Importantly, SnO$_2$/TCMH films demonstrated attenuated susceptibility to photo-induced trap formation, with spatial correlation showing trap activity primarily confined to the irradiated area (**Figure 3a (ii)**), whereas similar profile became apparent only after 2 minutes of recovery in TCMH only films. The dark zone regained its emission within 15 minutes (**Figure 3a (iii), Figure S8 a (iii)**), indicating faster healing than pristine. In contrast, ZnO/TCMH heterostructures exhibited high-CLIM correlation with significantly larger radial spread, extending up to ~10 µm from photodamaged center, with slow defect annihilation in the periphery (**Figure 3b (ii–iii)**). For PEDOT:PSS/TCMH, the spatial defect activity after photodamage mirrored the pristine film, displaying smaller correlated 'donut-shaped' regions with a less severely damaged center (**Figure 3c (ii-iii)**). On the other hand, PTAA/TCMH films displayed a distinct response, with suppressed emissivity and fluctuations at the center immediately after damage, whereas recovery induced formation and growth of high CLIM correlation domains both at the center and periphery (**Figure 3d (ii-iii)**), pointing to interface-facilitated defect redistribution.

Remarkably, beyond modulating local trap dynamics during healing, hetero-structured films revealed striking shifts in overall spatial CLIM correlation patterns before and immidiately after the photodamage (**Figure 3(a-d) (i-ii)**), a level of variation barely seen in pristine films (**Figure 1b (i, v)**). Specifically, films interfaced with SnO$_2$ and PEDOT:PSS exhibited overall decreased spatial CLIM correlation values, while ZnO and PTAA-based heterostructures showed a slightly increased defect activity following photodamage. These differences gradually dissipated over extended healing periods (~ 24 hours). These observations demonstrate interfaces as active modulators of defect properties, likely through carrier extraction and ion migration, which influence spatial defect densities and dynamics.

**Table 1.** Summary of fitting parameters obtained from fitting normalized PL intensity curves during dark healing with a bi-exponential recovery function: $y = y_0 + A_1\left(1 - e^{\frac{-x}{t_1}}\right) + A_2(1 - e^{\frac{-x}{t_2}})$. Intensities (I) were normalized relative to the pre-damaged baseline intensity ($I_0$).

| Samples | $y_0$ (= $I/I_0$) | Norm. $A_1$ | $t_1$ (/min) | Norm. $A_2$ | $t_2$ (/min) |
|---|---|---|---|---|---|
| **TCMH** | 0.18 | 0.357 | 44.79 | 0.642 | 140.45 |
| **SnO$_2$/TCMH** | 0.26 | 0.138 | 3.91 | 0.861 | 168.72 |
| **ZnO/TCMH** | 0.20 | 0.734 | 34.91 | 0.265 | 164.69 |
| **PEDOT:PSS /TCMH** | 0.10 | 0.418 | 7.84 | 0.581 | 133.06 |
| **PTAA /TCMH** | 0.19 | 0.526 | 9.79 | 0.473 | 150.65 |

Moreover, to compare recovery kinetics amongst pristine and heterostructure films, we analyzed normalized PL intensities ($I/I_0$) of photodamaged regions over the recovery period. Here, $I_0$ denotes pre-damage baseline intensity and $I$ corresponds to PL at specified intervals after photodamage. The analysis revealed that while pristine TCMH film experienced a ~ 80% decrease in PL immediately following photodamage (**Figure 3e**), heterostructures showed modest quenching variations (**Table 1**). Bi-exponential fits to recovery curves identified two characteristic regimes – fast ($t_1$), and gradual (slow, $t_2$). Interestingly, incorporation of CTLs significantly accelerated the initial $t_1$ component; specifically, films with SnO$_2$, PEDOT:PSS, and PTAA exhibited 4 to 10 times faster (under 10 min) than pristine films (**Table 1**). In contrast, recovery in ZnO interfaced samples progressed more slowly, akin to pure TCMH. The amplitudes of PL fluctuations returned to their initial values across all samples within $t_1$ phase. The slow recovery component ($t_2$) remained comparable among pristine and all heterostructures, indicating that CTLs mainly facilitate rapid initial defect passivation or carrier recombination restoration; however, they have little influence on the slower recovery process.

**Modulation of local energetic landscape during self-healing.** Building on prior characterizations of defect dynamics and SH, we used spatially resolved spectral imaging to probe energetic changes associated with defect formation and annihilation (**Figure 4**, **Figure S11**). Pristine TCMH films exhibited energetically homogeneous emission profile, with consistent peak positions and spectral shapes across the entire imaged area (**Figure 4a**). Upon localized photodamage, however, a pronounced spatial redistribution of emission energy emerged (**Figure 4b**). Specifically, the irradiated region showed a red shift in PL peak position of 10 meV, while the surrounding periphery (~ 5 μm away from the irradiated area) exhibited a progressive blue shift of 17 meV. Importantly, spectral shape remained largely unaltered, before and after photodamage, as demonstrated in PL spectra from two representative ROIs, ROI-1 (centre) and ROI-2 (periphery) (**Figure 4c**). This spatially inverse and reversible spectral shift pattern was traced, with emission energy gradually reverted to their initial values over the course of two hours of healing (**Figure 4d**).

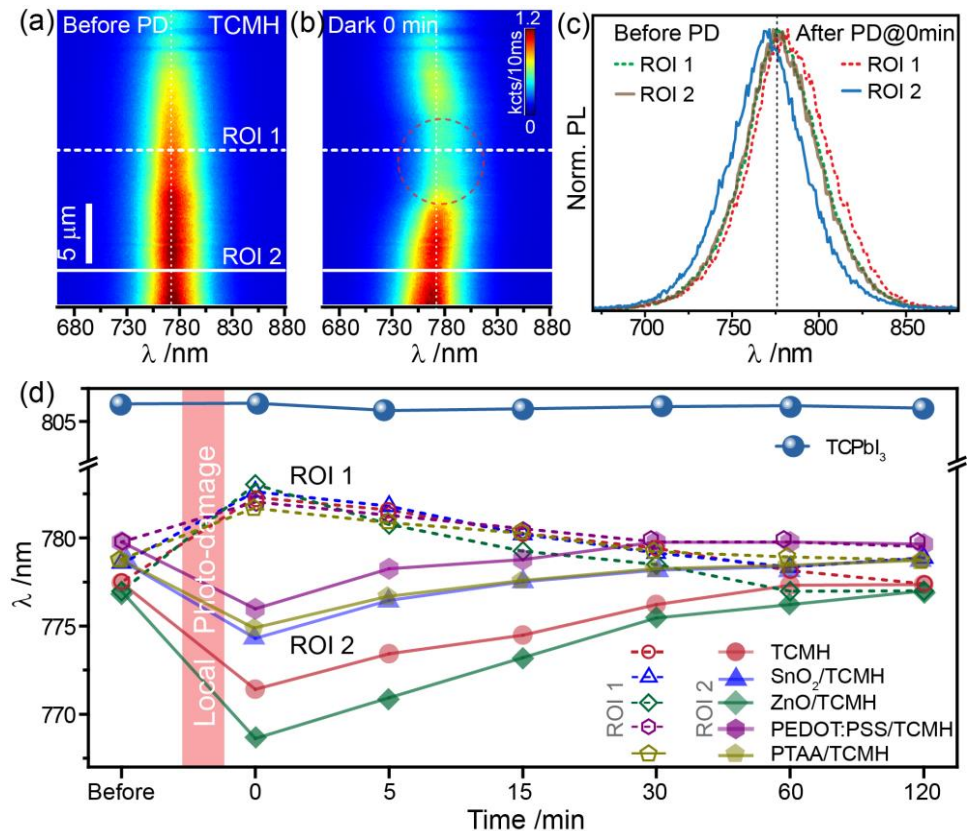

**Figure 4.** PL spectral images of TCMH film (a) prior, and (b) immediately after photodamage (0 min). (c) Representative PL spectra from ROI-1 (center) and ROI-2 (periphery, ~ 10 μm away) before and after photodamage, illustrating a blue shift at ROI-1 and red shift at ROI-2 relative to pristine film spectra. (d) Temporal evolution of spectral peak wavelengths of ROI-1 and ROI-2 during recovery of both pristine and hetero-structured TCMH films. Change in PL peak positions at the center of the TCPI$_3$ film is included for comparison.

These spectral modulations likely arise from halide segregation and/or photo-induced structural distortions by localized damage. To discern the origin(s), we performed analogous experiments on triple-cation iodide-only perovskite (TCPI$_3$) film (lacks bromide), which exhibited negligible spectral shifts under identical conditions (**Figure 4d**, **Figure S12**). This contrast strongly implicates halide ion migration and redistribution leading to the formation of bromide-rich (blue-shifted) and iodide-rich (red-shifted) regions, as the primary origin of the observed spectral heterogeneity in TCMH films.[34] Comparable spectral responses were observed in all TCMH heterostructures, though the extent and temporal progression of spectral shifts were CTL-dependent (**Figure S11**). A significant portion of spectral recovery occurred within the initial 15 minutes of SH; however, complete restoration of the electronic energy landscape required over an hour (**Figure 4d**). Amongst all heterostructures investigated, ZnO-based films showed the largest blue-shift (~ 18 meV), indicating greater disruption of the energetic landscape. Films with SnO$_2$, PEDOT:PSS, and PTAA exhibited moderate spectral shifts, emphasizing the critical role of interfacial chemistry in regulating both the severity of energetic disruption and restoration kinetics in these TCMH perovskite systems.

**Discussion**

Our observations establish that SH in TCMH films and their heterostructures is a multiscale process governed by transient defects, ion redistribution, and interfacial interactions. Localized photodamage generates metastable NR traps,[11,18] that quench light emission and disrupt carrier dynamics on the mesoscale. These traps likely originate from mobile defect accumulating at grain boundaries,[7,49] and/or from trapping of excess carriers in intra-band energy states.[11] Elevated local carrier densities create internal electric fields that lower ion migration barriers,[41,50] driving spatially patterned halide redistribution, leading to an iodide-rich irradiated center, surrounded by a bromide-enriched (or fewer Br⁻ vacancies) periphery. This behaviour likely stems from weaker Br⁻ bonding near anion vacancies, which enable vacancy-mediated outward diffusion during irradiation.[51] These findings align well with previous studies on mixed halide perovskites.[52,53] Despite these photophysical changes, no morphological degradation was observed (**Figure S13**), which reaffirms that photo-darkening arises from transient defects and ionic dynamics. Ion migration consequently propagates defect evolution into peripheral regions, reshaping recombination landscapes, which alter spatial correlation and clusters. Upon cessation of irradiation, halides slowly redistribute to re-establish equilibrium by de-trapping and de-activating excess defects. These results underscore the crucial role of mixed-halide mediated ionic transport in governing the dynamic emissive behavior and recovery of TCMH films.

The observed two recovery regimes of both TCMH and hetero-structured films can be interpreted as - $t_1$ (~ tens of minutes) corresponds to annihilation of light-induced metastable traps, while $t_2$ (~ hours) involves ionic and lattice rearrangements.[54] It is important to note that heterostructures, unlike pristine films, exhibited significant changes in overall spatial CLIM correlation even at distant sites, before and after the photodamage (**Figure 3**), and possess shorter $t_1$ than TCMH on glass (**Table 1**). We speculate that conductive CTL interfaces possibly help in efficient extraction and delocalization of excess carriers from the irradiated zone, thereby accelerating the process, an effect absent in case of an insulating glass substrate. It is likely that dispersed charges relax across broader interface, and/or interact with interfacial traps, thereby significantly modulating defect dynamics over large distances. The halide migration depends strongly on defect density, vacancies, and carrier concentration,[55–57] which is further modulated by interfaces. $SnO_2$ interface suppresses defect formation, limits halide migration,[58] and accelerates recovery, evident from reduced PL quenching, minimal spectral shift, and short $t_1$. In contrast, ZnO exacerbates interface trap activity promoting ion migration,[50] which caused pronounced spectral distortion. Among HTLs, PEDOT:PSS introduces additional traps due to its acidic nature, intensifying PL fluctuations and quenching, whereas PTAA suppresses both trap activity and halide migration.[20,59] Increased hole concentrations have been shown to accelerate halide segregation,[60,61] which rationalizes relatively less severe spectral shifts in HTL-based heterostructures (**Figure 4d**). Collectively, our findings emphasize that rational interface engineering aimed to minimizing defect density and impeding ion migration provides a compelling approach to improve the durability of perovskite photovoltaic devices by facilitating efficient defect self-healing.

**Conclusions and future perspectives**

We demonstrated a spatially resolved study of defect dynamics, ionic migration, and SH in TCMH perovskite films and their CTL heterostructures. Employing advanced PL imaging and analysis methods, we showed that SH in TCMH based films emerges from the interplay of metastable NR traps, halide ion redistribution, and ease of charge delocalization though CTLs. Our findings reveal that microscopic scale defect populations and spatially correlated emission domains evolve dynamically during photodamage and subsequent dark recovery, governed by distinct fast electronic and slower ionic relaxation processes. Importantly, chemically stable and inert CTLs, such as $SnO_2$ and PTAA, suppress trap formation, restrict halide migration, and accelerate recovery, indicating the crucial role of interface chemistry in achieving long-term device stability.

Our results emphasize that interface engineering is a powerful strategy to tailor defect landscapes and regulate ion transport. Further detailed studies are needed using atomic-scale, *in situ* high resolution probes to understand interface-defect interactions , and how conductive interfaces facilitate faster photo-induced defect healing. Complementary compositional approaches, such as targeted doping or passivation, when combined with interface engineering, hold promise to further suppress detrimental ion migration and fully realize the potential of perovskite photovoltaics and light-emitting devices with enhanced durability and performance.

**ASSOCIATED CONTENTS**

**Supporting Information**

The Supporting Information is available free of charge at-

Supporting text includes details of experimental methods; structural and optical characterization; spatial PL imaging and spectroscopy; correlated Fluo-SEM; and Monte-Carlo simulation. Supporting figures comprise SEM, UV-Vis, XRD, PL fluctuation traces od representative ROIs, CLIM cluster areas at irrdiated center and periphery of TCMH films, simulated blinking traces at varies defect densities, histograms of PL intensity and CLIM corrlation for heterostructures, PL images and CLIM clusters of heterostructures, and spectral images of heterostructres and $TCPbI_3$ film (PDF).

**Supporting Movies**

PL movies of TCMH film (M1)

## Author Information

**Corresponding Author**

**Sudipta Seth** − Laboratory for Photochemistry and Spectroscopy, Division for Molecular Imaging and Photonics, Department of Chemistry, Katholieke Universiteit Leuven, Leuven 3001, Belgium; orcid.org/0000-0002-8666-4080; Email: sudipta.seth@kuleuven.be

**Johan Hofkens** − Laboratory for Photochemistry and Spectroscopy, Division for Molecular Imaging and Photonics, Department of Chemistry, Katholieke Universiteit Leuven, Leuven 3001, Belgium; Max Planck Institute for Polymer Research, Mainz 55128, Germany; orcid.org/0000-0002- 9101-0567; Email: johan.hofkens@kuleuven.be

**Authors**

**Tejmani Behera** - Laboratory for Photochemistry and Spectroscopy, Division for Molecular Imaging and Photonics, Department of Chemistry, Katholieke Universiteit Leuven, Leuven 3001, Belgium; orcid.org/ 0000-0003-2995-9293; Email: tejmanibehera718@gmail.com

**Boris Louis** − Laboratory for Photochemistry and Spectroscopy, Division for Molecular Imaging and Photonics, Department of Chemistry, Katholieke Universiteit Leuven, Leuven 3001, Belgium

**Lukas Paesen** - Laboratory for Photochemistry and Spectroscopy, Division for Molecular Imaging and Photonics, Department of Chemistry, Katholieke Universiteit Leuven, Leuven 3001, Belgium

**Roel Vanden Brande** - Laboratory for Photochemistry and Spectroscopy, Division for Molecular Imaging and Photonics, Department of Chemistry, Katholieke Universiteit Leuven, Leuven 3001, Belgium

**Koki Asano** - Department of Materials Science and Engineering, Tokyo Institute of Technology, Meguro-ku, Tokyo 152-8552, Japan

**Martin Vacha** - School of Materials and Chemical Technology, Institute of Science Tokyo, Tokyo 152-8552, Japan; orcid.org/0000-0002-5729- 9774

**Maarten B. J. Roeffaers** − Laboratory for Photochemistry and Spectroscopy, Division for Molecular Imaging and Photonics, Department of Chemistry, Katholieke Universiteit Leuven, Leuven 3001, Belgium; orcid.org/0000-0001-6582-6514

**Elke Debroye** − Laboratory for Photochemistry and Spectroscopy, Division for Molecular Imaging and Photonics, Department of Chemistry, Katholieke Universiteit Leuven, Leuven 3001, Belgium; orcid.org/0000-0003-1087-4759

**Author Contributions**

T.B. fabricated the films used in this study, collected and analyzed the data. B.L. provides the simulation code. T.B., S.S., and B.L. together interpret the results. T.B. wrote the initial draft and with SS, BL, MR, ED, and JH finalized the paper. LP helped in synthesis and partially involved in PL measurements. RVB performed and analysed the XRD measurements. All authors reviewed the final text before submission, helping to clarify and simplify the text.

**Notes**

The authors declare no competing financial interest.

**Acknowledgements**

S.S. acknowledges the support of Marie Skłodowska-Curie postdoctoral fellowship (No. 101151427, SPS Nano) from the European Union's Horizon Europe program, short stay abroad grant (K257023N) and travel grant (K147824N) from Research Foundation-Flanders (FWO). B.L. thanks FWO for his Junior Postdoctoral fellowship (12AGZ24N). E.D. acknowledges funding from the KU Leuven Internal Funds (grant numbers C14/23/090 and CELSA/23/018), the Research Foundation-Flanders (FWO, grant number G0AHQ25N) and the European Union (ERC Starting Grant, 101117274 X-PECT). J.H. acknowledges financial support from the Research Foundation-Flanders (FWO, Grant Nos. G098319N, G0F2322N, S002019N, VS06523N, G0AHQ25N) from the Flemish government through long-term structural funding Methusalem (CASAS2, Meth/15/04), and the MPIP as MPI fellow. M.V. acknowledges financial support by the JSPS KAKENHI grant number 24K01449 and by the JSPS KAKENHI grant number 23H04875 in Grant-in-Aid for Transformative Research Areas 'Materials Science of Meso-Hierarchy'. However, the views and opinions expressed are those of the authors only and do not necessarily reflect those of the European Union or European Research Council. Neither the European Union nor the granting authority can be held responsible for them.